\documentclass{aa}
\usepackage{graphicx}
\usepackage{txfonts}

\def\spose#1{\hbox to 0pt{#1\hss}}
\def\simless{\mathrel{\spose{\lower 3pt\hbox{$\mathchar"218$}}
        \raise 2.0pt\hbox{$\mathchar"13C$}}}
\def\simgreat{\mathrel{\spose{\lower 3pt\hbox{$\mathchar"218$}}
        \raise 2.0pt\hbox{$\mathchar"13E$}}}\def\lta{\mathrel{\spose{\lower 3pt\hbox{$\mathchar"218$}}
        \raise 2.0pt\hbox{$\mathchar"13C$}}}
\def\gta{\mathrel{\spose{\lower 3pt\hbox{$\mathchar"218$}}
        \raise 2.0pt\hbox{$\mathchar"13E$}}}
\def\msol{\rm M_\odot}
\begin{document}

\title{XMM-Newton observations of two black hole X-ray transients in quiescence}

\author{J.-M. Hameury\inst{1}, D. Barret\inst{2}, J.-P. Lasota\inst{3},
J.E. McClintock\inst{4}, K. Menou\inst{5}\thanks{Celerity
Foundation Fellow}, C. Motch\inst{1}, J.-F. Olive\inst{2}, N.
Webb\inst{2}}
\offprints{J.-M. Hameury,\\
\email{hameury@astro.u-strasbg.fr}}

\institute{
UMR 7550 du CNRS, Observatoire de Strasbourg, 11 rue de
           l'Universit\'e, F-67000 Strasbourg, France
\and Centre d'Etude Spatiale des Rayonnements, 9 Av. du Colonel
Roche, BP 4346, 31028 Toulouse Cedex 4, France \and UMR 7095 du
CNRS, Institut d'Astrophysique de Paris, 98bis Boulevard Arago,
75014 Paris, France \and Harvard-Smithsonian Center for
Astrophysics, 60 Garden Street, Cambridge, MA 02138, USA \and
Department of Astronomy, PO Box 3818, University of Virginia,
Charlottesville, VA 22937, USA }

\date{Received / Accepted}
\titlerunning{Soft X-ray transients in quiescence}
\authorrunning{Hameury et al.}

\abstract{We report on XMM-Newton observations of GRO J1655-40 and
GRS 1009-45, which are two black hole X-ray transients currently
in their quiescent phase. GRO J1655-40 was detected with a 0.5 -
10 keV luminosity of $5.9 \times 10^{31}$ erg s$^{-1}$. This
luminosity is comparable to a previous Chandra measurement, but
ten times lower than the 1996 ASCA value, most likely obtained
when the source was not yet in a true quiescent state.
Unfortunately, XMM-Newton failed to detect GRS 1009-45. A
stringent upper limit of $8.9 \times 10^{30}$ erg s$^{-1}$ was
derived by combining data from the EPIC-MOS and PN cameras.

The X-ray spectrum of GRO J1655-40 is very hard as it can be
fitted with a power law model of photon index $\sim 1.3\pm 0.4$.
Similarly hard spectra have been observed from other systems;
these rule out coronal emission from the secondary or disk flares
as the origin of the observed X-rays. On the other hand, our
observations are consistent with the predictions of the disc
instability model in the case that the accretion flow forms an
advection dominated accretion flow (ADAF) at distances less than a
fraction ($\sim$ 0.1 - 0.3) of the circularization radius. This
distance corresponds to the greatest extent of the ADAF that is
thought to be possible.
  } \maketitle

\begin{table*}
\caption{Main parameters of quiescent BH SXTs with known orbital
periods P$_{\rm orb}$. $M_1$ is the primary mass, and $n_{\rm H}$
the hydrogen column density, generally inferred from optical
extinction. The luminosity refers to the 0.5 - 10 keV energy
range; $\alpha$ is the power law photon index. The orbital periods
and masses have been taken from Narayan et al. (\cite{ngm02} and
references therein; data for GRO J1655-40 are from Beer \&
Podsiadlowski (\cite{bp02})). The dividing line separates somewhat
arbitrarily systems with short and long orbital periods.}
\begin{tabular}{lrccclcl}
\hline
             &$P_{\rm orb}$&D& $M_1$ & $n_{\rm H}$& $\log L_{\rm X}$  & $\alpha$ & Reference\\
             &  (hr)  & (kpc)  & (M$_\odot$)& (10$^{21}$ cm$^{-2}$) &  \\
\hline
\object{XTE J1118+480} &  4.1 & 1.8  & 7 $\pm$ 1   & 0.1 &       & & \\
\object{GRO J0422+32}  &  5.1 & 2.6  & 10 $\pm$ 5  & 1.6 & 30.9  & -- & Chandra (3)\\
\object{GRS 1009-45}   &  6.9 & \ \ \ 5.0$^{[*]}$  & 4.2$\pm$ 0.6& 1.1 &$<$30.9& & XMM-Newton -- this paper\\
\object{A0620-00}      &  7.8 & 1.0  & 10 $\pm$ 5  & 1.9 & 30.5  & 2.07$^{+0.28}_{-0.19}$& Chandra (1)\\
                       &      &      &             & 0.16& 30.7  &  & ROSAT (5)\\
\object{GS 2000+25}    &  8.3 & 2.7  & 10 $\pm$ 4  & 11  & 30.4  & -- & Chandra (3)\\
\object{XTE J1859+226} & 9.2  &      & 10 $\pm$ 3  &     &       & &\\
\object{Nova Mus 1991} & 10.4 & 5.0  & 7 $\pm$ 3   & 2.6 &   31.6& 1.6$\pm$0.4&XMM-Newton (2)\\
\object{Nova Oph 1977} & 12.5 & 8.0  & 4.9$\pm$ 1.3& 2.8 &$<$33.0& & ROSAT (4)\\
\hline
\object{4U1543-47}     & 27.0 & 9.0  & $5 \pm$ 2.5 & 3.5 &$<$31.5& & Chandra (3)\\
\object{XTE J1550-564} & 37.0 & 6.0  & $>$ 7.4     & 3.9   & 33.1  & 1.35$\pm$0.25&Chandra (1)\\
\object{GRO J1655-40}  & 62.9 & 3.2  & 5.4 $\pm$ 0.3   & 6.7 & 31.5  & 1.47$\pm$0.4&Chandra (1)\\
                       &      &      &             &     & 31.8  & 1.30$\pm$0.40& XMM-Newton -- this paper\\
                       &      &      &             & $<$3& 33.1  & 0.7$^{+0.21}_{-0.4}$& ASCA (6)\\
\object{V4641 Sgr}     & 67.6 &10.0  & 10.2$\pm$1.5& 0.5 &       & & \\
\object{V404 Cyg}      &155.3 & 3.5  & 12 $\pm$ 2  & 5.4 & 33.6 & 1.55 $\pm$ 0.07 & Chandra (1) \\
                       &      &      &                   & 11$^{+3}_{-4}$ & 33.1 & 2.1$^{+0.3}_{-0.3}$ & ASCA (7) \\
                       &      &      &                   & 10             & 33.0 & 1.9$^{+0.6}_{-0.3}$ & Beppo-SAX (8) \\
\hline
\end{tabular}
\medskip

\noindent References: (1) Kong et al. (\cite{kmg02}); (2) Sutaria
et al. (\cite{skc02}); (3) Garcia et al. (\cite{gmn01}); (4)
Verbunt et al. (\cite{vbj94}); (5) Narayan et al. (\cite{nmy96});
(6) Asai et al. (\cite{adh98}); (7) Narayan et al. (\cite{nbm97});
(8) Campana et al.
(\cite{cps01}); \\
 {[*]}from Barret et al. (\cite{betal00}) \label{tab:sxt}
\end{table*}

\section{Introduction}

Soft X-ray transients -- SXTs (sometimes called X-ray novae) are
semi--detached binaries in which the accreting (primary) star is a
black hole (BH) or a neutron star, and the mass--losing secondary
is usually a late type star. These systems typically brighten in
X-rays by as much as 10$^7$ in a week and then decay back into
quiescence over the course of a year. The maximum outburst
luminosities $L_{\rm max}$ seen in SXTs are typically
$\sim(0.2-1)$ of the Eddington luminosity $L_{\rm Edd}$.
Successive outbursts are usually separated by years to decades of
quiescence (see e.g. Tanaka \& Shibazaki \cite{ts96}; Chen et al.
\cite{csl97}, for reviews).

Quiescent states of SXTs are very interesting for at least two
reasons. First, they provide the strongest evidence now available
for the existence of stellar--mass black holes. The detection of
the secondary in quiescence allows a determination of the mass
function of the binary system, which is an absolute lower limit on
the primary mass $M_1$. The mass function exceeds 3 M$_\odot$ in
eight systems; in five others, constraints on the inclination
angle of the system and the secondary mass result in values for
$M_1$ that are in the range 5 -- 10 M$_\odot$ (Narayan et al.
\cite{ngm02}). Thus there are now thirteen SXTs with primary
masses that exceed the maximum stable mass of a neutron star
($\simless 3$ M$_\odot$).

The second, more mundane reason why quiescent states are
interesting is that quiescence provides a strong test of
outburst--cycle models. SXT outburst cycles are well described by
the disc instability model (Dubus et al. \cite{dhl01}; see Lasota
\cite{l01} for a complete review of the disc instability model).
In quiescence, the disc is non--steady (a property too often
forgotten by too many authors). Its temperature and viscosity are
low, and the disc is unable to transport all of the mass supplied
by the secondary to the primary; the mass of the disc slowly
builds up and the temperature rises finally to the hydrogen
ionization temperature. At this moment, the disc becomes thermally
and viscously unstable due to strong opacity variations.
Propagating heat fronts bring the entire disc into a hot state; in
this outburst state, the mass transfer rate is large, and the disc
empties until it cannot sustain this regime any longer; it then
returns to quiescence.

Even though the disc instability paradigm is widely accepted and
its particular realizations are often successful (requiring
sometimes additional assumptions, see e.g. Esin et al.
\cite{elh00}), the physics of the quiescent state is still rather
controversial. As a prime example, the origin of ``viscosity" in
this state is not really known (see e.g. Gammie \& Menou,
\cite{gm98}; Menou \cite{m02}; Lasota \cite{l02}), and according
to the simplest version of the disc-instability model (the version
originally used to explain dwarf-nova outbursts; Lasota
\cite{l01}) the very long SXT recurrence times require unusually
low values of the viscosity parameter $\alpha$. However, Dubus et
al. (\cite{dhl01}) showed that the combined effects of disc
irradiation during outbursts (King \& Ritter \cite{kr98}) and disc
truncation during quiescence (Menou et al. \cite{mhln00}) result
quite effortlessly in long recurrence times for standard values of
the viscosity parameter, making the disc instability model for
SXTs a working and testable hypothesis.

Narayan et al. (\cite{nmy96}; see also Lasota et al. \cite{lny96}
and Narayan et al. \cite{nbm97}) noticed that spectra of quiescent
SXTs cannot be produced by geometrically thin, optically thick
accretion discs and suggested that the inner flow in such systems
forms an advection dominated accretion flow ADAF. In any case,
according to the disc instability model a disc extending down to
the neutron star surface or the last stable orbit around a black
hole, can be in a cold, neutral state everywhere only for
vanishingly low accretion rates close to the central object
(Lasota \cite{l96}): the maximum accretion rate onto the compact
object would be $\sim 4000 \left(M_1/\msol\right)^{1.77} (r_{\rm
in}/r_{\rm s})^{2.65}$ g s$^{-1}$, where $r_{\rm s}$ is the
Schwarzschild radius (Hameury et al. \cite{hmd98}), much too low
to produce the X-ray luminosity observed in quiescent SXTs. On the
other hand the ADAF model reproduces well the observed
luminosities, spectra (Narayan et al. \cite{nbm97}; Quataert \&
Narayan \cite{qn99}) and observed delays between the rises to
outburst in optical and X-rays (Hameury et al. \cite{hlmn97}).

As first pointed out by Narayan et al. (\cite{ngm97}), the
presence of the radiatively inefficient ADAFs in quiescent SXTs
allows one to compare black holes with compact bodies endowed with
material surfaces. Because bodies such as neutron stars must
re-emit the heat left over in the accretion flow accumulating at
their surface they should be brighter than black holes accreting
at the same rate since in this case the residual energy is lost
forever past the event horizon. X-ray observations of quiescent
SXTs showed that this is indeed the case (Narayan et al.
\cite{ngm97}; Menou et al. \cite{men99}; Garcia et al.
\cite{gmn01}). However, the ADAF origin of X-rays in quiescent
SXTs has been questioned by several authors (see e.g. Bildsten \&
Rutledge \cite{br01}; Nayakshin \& Svensson, \cite{ns01}). High
quality X-ray observations can solve this controversy.

In this paper, we report on XMM-Newton observations of \object{GRO
J1655-50} (AO-1 GTO) and \object{GRS 1009-45} (AO-1 GO).We briefly
describe previous observations of these transients in section 2;
we present our observations in section 3, and we discuss their
implications in section 4.

\section{Earlier observations}

Observations of quiescent SXTs prior to XMM-Newton and Chandra
have yielded only marginal detections or upper limits, with very
poor constraints on the spectral shape. Among the 13 BHSXTs with
known orbital periods, \object{A 0620-00} was the only one to be
detected by ROSAT (McClintock et al., \cite{mhr95}); ASCA detected
\object{GRO J1655-40} (Asai et al. \cite{adh98}; Ueda et al.
\cite{uit98}) and \object{V404 Cyg} (Narayan et al. \cite{nbm97}).
Finally, among the 5 BHSXTs observed by BeppoSAX (Campana et al.
\cite{cps01}), only \object{V404 Cyg} has been detected with a
1-10 keV unabsorbed luminosity of $\sim 10^{33}$ erg s$^{-1}$.

XMM-Newton and Chandra have already qualitatively changed this
situation, as can be clearly seen from Table \ref{tab:sxt}, which
gives spectral data for most of the 13 SXTs with known orbital
period, and with primary masses exceeding 3 M$_\odot$.

It is now evident that many quiescent SXTs have been detected with
luminosities in the range 10$^{31}$ -- 10$^{33}$ erg s$^{-1}$, the
faintest being those with the shortest orbital periods. It is also
clear that SXTs have very hard spectra with values of the power
law spectral index that are often less than 2. Or equivalently,
their spectra can be fitted by a bremsstrahlung model with
temperature larger than 10 keV (Kong et al. \cite{kmg02}).

\section{XMM-Newton observations and analysis}

\subsection{\object{GRO J1655-40}}

GRO J1655-40 was observed by XMM-Newton on August 30, 2001 with a
40 ksec exposure time. Due to a strong solar flare, only 19 ks
could be used; yet, GRO J1655-40 was clearly detected as one of
the brightest field sources, at a position coincident with that of
the optical counterpart (within 0.3\arcsec), with a total source
count rate as determined by the automatic pipeline analysis of
1.09 10$^{-2}$ $\pm$ 1.8 10$^{-3}$.

\begin{figure}
\resizebox{\hsize}{!}{\includegraphics[angle=-90]{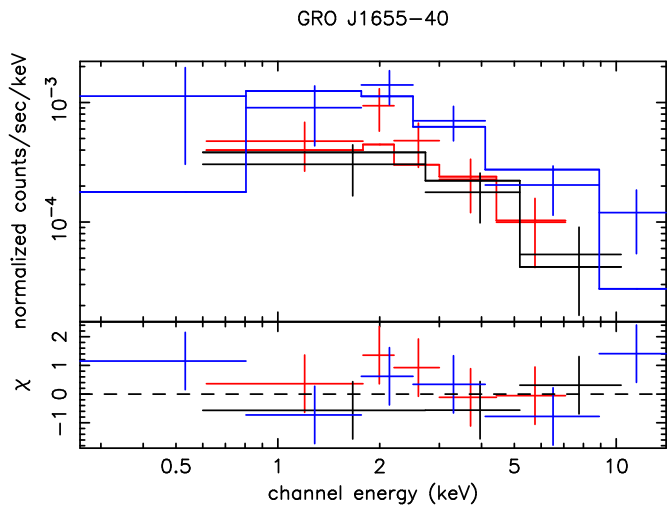}}
\caption{Top: XMM-Newton spectrum of GRO J1655-40, fitted by a
power law. Bottom: residuals after model substraction from the
data, in units of 1 $\sigma$. The pn data are in blue, the MOS
data in black and red.} \label{fig:1655_spec}
\end{figure}

\begin{figure}
\resizebox{\hsize}{!}{\includegraphics[angle=-90]{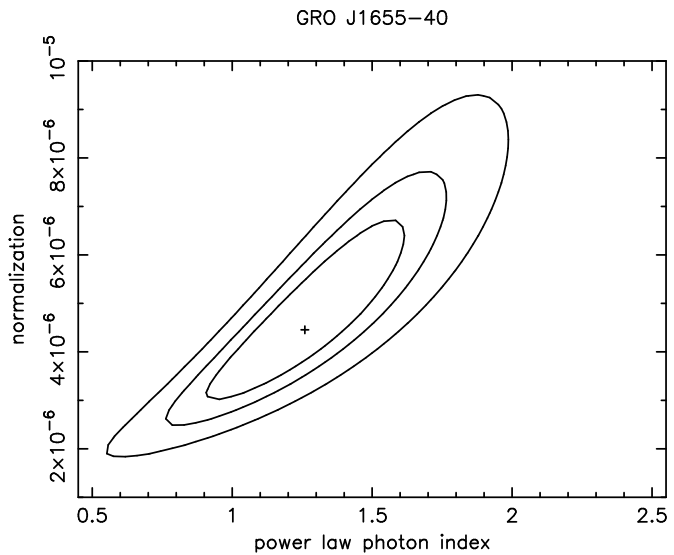}}
\caption{Confidence level at 68, 90 and 99 \% of the fit given in
Fig. \ref{fig:1655_spec}. $n_{\rm H}$ is kept fixed at 6.7
10$^{21}$ cm$^{-2}$.} \label{fig:1655_cont}
\end{figure}

SAS version 5.3 was used to calculate the source and background
spectra obtained by extracting data from a circle of approximately
20{\arcsec} radius centered on the nominal source position, and
from an annulus with inner and outer radii $\sim$ 20{\arcsec} and
$\sim$ 100\arcsec respectively. 74, 73 and 260 counts were
detected in the 20{\arcsec} radius circle in the MOS1, MOS2 and PN
cameras respectively; 618 (MOS1), 403 (MOS2) and 1225 (PN) counts
were detected in the larger area used to determine the background
rate. The spectra were analyzed with Xspec v11. Due to the small
number of counts, in particular in the MOS cameras, the
C-statistics (Cash, \cite{c79}) was used to determine the best
fits and errors; channels were grouped in such a way that there
were at least 5 counts in each bin. The three cameras, MOS1, MOS2,
and PN were used simultaneously to determine the best fit
parameters.

\begin{table*}
\caption{Best fitting spectral parameters for GRO J1655-40.
$F_{0.5-10}$ is the 0.5 - 10 keV unabsorbed flux. The goodness is
the probability that simulated data give a better C statistic than
real data.}
\begin{tabular}{lccccccc}
\hline
Model          &$n_{\rm H}$         & $\alpha$ & kT & C statistic & dof & goodness & $F_{0.5-10}$ \\
               &(10$^{21}$ cm$^{-2}$&          & (keV)&    &  & & 10$^{-14}$ (erg cm$^{-2}$s$^{-1}$) \\
\hline power law      &10.0$^{+1.2}_{-0.7}$ &
1.54$^{+1.02}_{-0.72}$& &
63.8 & 72 & 0.12 & 5.3 \\
power law      &6.7 (fixed) & 1.30$^{+0.34}_{-0.41}$ & & 64.5 & 73 & 0.11 & 4.8\\
bremsstrahlung &6.7 (fixed) & & $>$ 9 & 64.5 & 73 & 0.12 & 4.1 \\
\hline
\end{tabular}
\label{tab:fits}
\end{table*}

Table \ref{tab:fits} summarizes our results. A power law is a good
fit to the results, yielding column density $n_{\rm H} =
1.0^{+1.2}_{-0.7} 10^{22}$ cm$^{-2}$, a power law index $\alpha =
1.54^{+1.02}_{-0.72}$, and normalization $K$ = 7.1 10$^{-6}$,
poorly constrained because of the large error in $\alpha$. If one
uses a hydrogen column density fixed to the value of $n_{\rm H}$ =
6.7 $10^{21}$ cm$^{-2}$ determined by optical observations (Hynes
et al. \cite{hhs98}), one obtains $\alpha = 1.30^{+0.34}_{-0.42}$,
and $K$ = 4.8$^{+2.0}_{-1.9}$ 10$^{-6}$. All errors given here are
90\% confidence limits. Figure \ref{fig:1655_spec} shows the EPIC
spectrum for all three instruments, as well as the best fit and
residuals, in units of 1 sigma. Note that these residuals,
although giving a rough estimate of the quality of the fit, cannot
be used to determine the goodness-of-fit by computing a $\chi^2$,
since the C-stat has been used. Note also that, for the sake of
clarity, data in Fig. \ref{fig:1655_spec} have been rebinned, so
that more data points were used for performing the spectral
fitting. Figure \ref{fig:1655_cont} gives the 68.3, 90 and 99\%
confidence contours obtained for the power law index and
normalization.

The total absorbed flux is then 3.96$^{+0.66}_{-0.83}$ 10$^{-14}$
erg cm$^{-2}$ s$^{-1}$ in the 0.5 - 10 keV range (90\% confidence
error bars) , corresponding to an unabsorbed flux of
4.81$^{+0.80}_{-1.00}$ 10$^{-14}$ erg cm$^{-2}$ s$^{-1}$ and
unabsorbed luminosity of 5.9 10$^{31}$ erg s$^{-1}$ for a distance
of 3.2 kpc.

We also fitted the spectrum with a thermal bremsstrahlung; as
expected for a hard spectrum with such a photon index, the fit is
good, with a formal best fit value of the temperature $kT$ = 110
keV, indeed unphysical, and very poorly constrained; the 90\%
confidence level lower limit on $kT$ is 9 keV, and the 99\%
confidence level lower limit is 4.1 keV.

\begin{figure}
\resizebox{\hsize}{!}{\includegraphics[angle=-90]{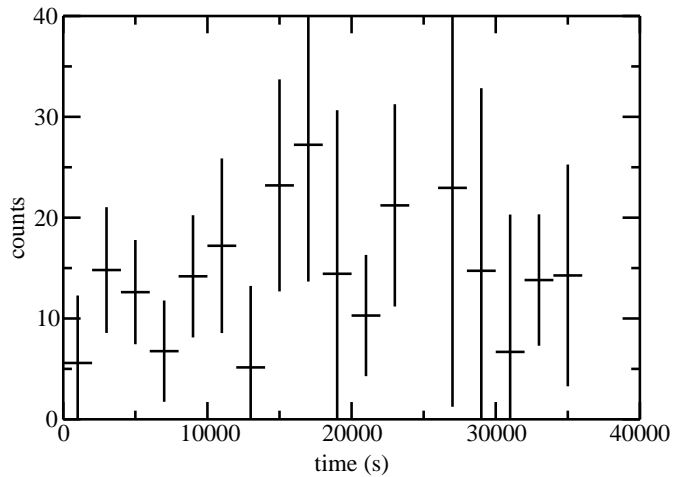}}
\caption{Net count rate per 2000 s bins for GRO J1655-40 in the PN
camera. Large errors are produced by the strong solar flare.}
\label{fig:1655_rate}
\end{figure}

A search for variability yielded negative results for significant
variations on timescales of hours. Figure \ref{fig:1655_rate}
shows the source intensity as a function of time, with 2000 s time
bins; the data are consistent with a constant source. The data
quality is quite poor during the solar flare that occurs after the
first 10 ks, and lasts about 25 ks, with shorts lulls.

\subsection{\object{GRS 1009-45}}

GRS 1009-45 was observed on May 30, 2002, during a 20 ksec
exposure. The source was not detected by the automatic pipeline
processing system. Assuming a power law spectrum with photon index
2, and a hydrogen column density $n_{\rm H}$ = 1.1 10$^{21}$
cm$^{-2}$, as given by optical observations (della Valle et al.
\cite{dbc97}), and analyzing simultaneously all three cameras
together, we can set an upper limit for the 0.5 - 10 keV
unabsorbed flux of 3 10$^{-15}$ erg cm$^{-2}$ s$^{-1}$ (95\%
confidence), corresponding to an upper limit on the luminosity of
8.9 10$^{30}$ erg s$^{-1}$ for a distance of 5 kpc. To determine
this upper limit, we fixed the photon index and column density in
Xspec and calculated the goodness-of-fit for various normalization
factors, with the $\chi^2$ statistics (using the C-statistics
yields an almost identical upper limit, as expected). The data are
compatible with a zero flux (we obtain a goodness-of-fit of 63\%);
we can set an upper limit on the flux of 3 10$^{-15}$ erg
cm$^{-2}$ s$^{-1}$, for which the goodness-of-fit is 0.95, i.e. in
only 5 \% of the simulated cases did we find a C statistic value
worse than the value actually obtained. This upper limit is almost
independent of the assumed photon power law index.

\section{Discussion}

It has been suggested (Bildsten \& Rutledge \cite{br01}) that
X-rays detected from quiescent black-hole SXTs are not produced by
accretion, but are instead produced in the secondaries' hot
stellar coronae. There is now growing evidence that this is not
the case because the X-ray flux is too large (especially the $L_X$
relative to the bolometric luminosity of the secondary) and/or
because the spectrum is too hard to be emitted by a stellar corona
(Kong et al. \cite{kmg02}). As we shall see, the flaring disc
model also faces similar difficulties, whereas the ADAF
explanations can still account for the observations.

\subsection{Spectra}

The spectra of all black hole transients appear to be quite hard,
\object{GRO J1655-40} being the hardest, and \object{A0620-00}
having the steepest power law index; fits using bremsstrahlung
models also lead to very high temperatures. This confirms previous
findings (Kong et al. \cite{kmg02}) that coronal emission from the
secondary is ruled out. Einstein surveys of late type stars have
shown that the coronal temperatures of isolated stars do not
significantly exceed 1 keV, and reach at most a few keV in the
case of giant RS CVn stars (Schmitt \cite{scs90}). (Coronal
emission is additionally ruled out by the high X-ray luminosities;
see below.)

It is also most unlikely that such spectra could be produced by
emission from magnetic loops above the disc (see e.g. Nayakshin \&
Svensson \cite{ns01}) that would very probably result in spectra
similar to those of stellar coronae. Moreover, this model implies
that the accretion disc of quiescent dwarf novae should emit a
comparable soft X-ray flux; these systems are indeed X-ray bright,
with luminosities in the range 10$^{30}$ - 10$^{32}$ erg s$^{-1}$,
but they have deep eclipses indicating that the X-ray emitting
region is very close to the white dwarf; the residual X-ray
luminosity during eclipses is very low (e.g. of the order of 9
10$^{28}$ erg s$^{-1}$ for \object{OY Car}, Ramsay et al.
\cite{rpm01}).

\subsection{Luminosities}

\begin{figure}
\resizebox{\hsize}{!}{\includegraphics[angle=-90]{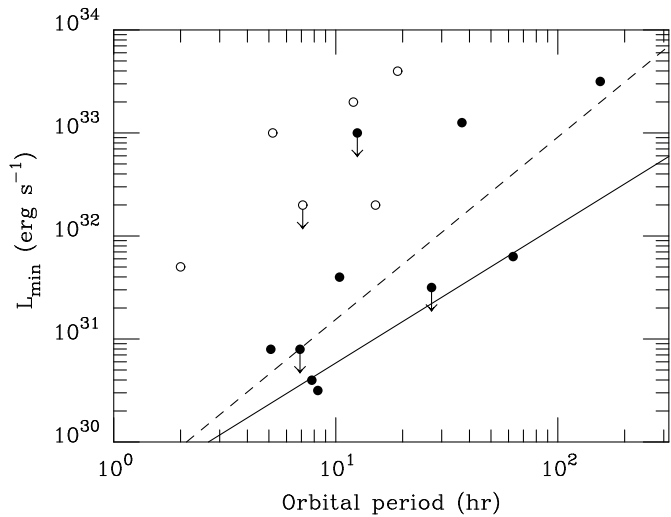}}
\caption{Quiescent luminosities of SXTs containing black holes
(filled circles: Chandra and XMM-Newton data from table
\ref{tab:sxt}), or a neutron star (open circles: data from Narayan
et al. (\cite{ngm02}), except for SAX J1808.4-3658 for which we
used XMM-Newton data of Campana et al. (\cite{csg02}), who found
the source in deeper quiescence than Wijnands (\cite{w02})). Also
shown are the maximum coronal X-ray luminosity (the "FGM-limit",
Flemming et al. \cite{fgm89}) for a 1 M$_\odot$ secondary (solid
line), and (dashed line) the flux predicted by Lasota
(\cite{l00})} \label{fig:nsbh}
\end{figure}

Figure \ref{fig:nsbh} shows the luminosities of SXTs containing
black holes compared to those in which the accreting object is a
neutron star. We plot the quiescent luminosity (in erg s$^{-1}$)
as a function of the orbital period of the binary. Here, the data
for BH systems have been taken from Table 1; when several
measurements were available, we selected Chandra or XMM-Newton
values (for GRO J1655-40, the XMM-Newton value given in this paper
is used). The NS data are from Narayan et al. (\cite{ngm02}). Our
result confirms the well established fact that systems containing
black holes are, for a given orbital period, much dimmer than
those containing a neutron star (i.e. in which X-ray bursts have
been detected), a property predicted by models in which the inner
disc is truncated and replaced by an ADAF close to the black hole
(Menou et al. \cite{men99}).

The $L_{\rm X} - P_{\rm orb}$ plot was proposed by Lasota \&
Hameury (\cite{lh98}) to assure, when one compares black-hole
systems with those containing neutron stars, that both types of
systems have comparable accretion rates {\sl at the outer limit of
the ADAF}. Sutaria et al. (\cite{skc02}) pointed out recently that
for various reasons one should expect a ``greater spread" of $\dot
M$ (the {\sl mass-transfer rate}) at a given $P_{\rm orb}$. This
is obviously true. However, if SXT outbursts are due to a
dwarf-nova-type disc instability, then in quiescence the accretion
rate at the truncation radius depends only weakly on the
mass-transfer rate (e.g. Menou et al. \cite{mnl99}, Lasota
\cite{l00}, Dubus et al. \cite{dhl01}; but see also Menou
\cite{m02} who discusses uncertainties linked to our poor
understanding of viscosity in the quiescent state and a possible
alternative solution). Indeed, the accretion rate at the
transition radius can be written as (Lasota \cite{l00})
\begin{equation}
\dot M_{\rm tr} \approx 2.4 \times 10^{15} \left(\frac{M}{7
M_{\odot}}\right)^{1.77}
  \left(\frac{R_{\rm tr}}{10^4R_{\rm G}}\right)^{2.65} \ {\rm g} \
{\rm s}^{-1}.
\label{accr}
\end{equation}

In addition, as discussed in Lasota (\cite{l00}) the prescription
for the transition radius proposed by Menou et al. (\cite{mnl99}),
i.e.
\begin{equation}
R_{\rm tr}=f_t \ R_{\rm circ} ,\qquad f_t<0.48 \label{rtr},
\end{equation}
where the circularization radius $R_{\rm circ}$ can be
approximated by (Frank et al. \cite{fkr02})
\begin{equation}
\frac{R_{\rm circ}}{a} = (1+q)[0.5-0.227 \times \log q]^4,\\
\label{eq:rcirc}
\end{equation}
transforms Eq. (\ref{accr}) into
\begin{equation}
\dot M_{\rm ADAF} \approx 1.6 \times 10^{18} f_t^{2.65} P_{\rm
day}^{1.77} {\rm g \ s^{-1}}
\label{slope}
\end{equation}
where $\dot M_{\rm ADAF}$ is the rate at which matter enters the
ADAF from the cold quiescent, {\sl non-steady} disc. This rate is
independent of the primary's mass.

To obtain Eq. (\ref{accr}) one assumes that according to the disc
instability model the quiescent disc is non-steady and $R_{tr} \ll
R_{\rm out}$, where $R_{\rm out}$ is the outer radius of the
geometrically thin accretion disc. The pre-Chandra and
pre-XMM-Newton data (for three systems) satisfied a $L_X \sim
P_{\rm orb}^{1.77}$ relation (dashed line in Fig. \ref{fig:nsbh}),
the slope of which suggested that it might result from Eq.
(\ref{slope}) (Lasota \cite{l00}).

The new Chandra and XMM-Newton observations show a more
complicated picture. First, the quiescent X-ray luminosity of GRO
J1655-40 is now an order of magnitude lower than it was when
observed by ASCA. In fact the quiescent luminosities of all three
systems (A0620-00, GRO J1655-40 and V404 Cyg) used by Lasota
(\cite{l00}) to determine the $\sim$ 1.77 slope of the $L_{\rm X}
- P_{\rm orb}$ relation have been revised significantly (see Table
\ref{tab:sxt} and discussion below). In the case of A0620-00 and
V404 Cyg there is little doubt that they are in true quiescence.
Among the new sources, XTE J1550-564 was observed between two
outbursts (see Kong et al. \cite{kmg02} and references therein) so
that this system was very unlikely to be in its true quiescent
state during the Chandra observations.

The quiescent luminosity might still be described by Eq.
(\ref{slope}) but with $f_t$ depending on the system. Such a
conclusion is not unexpected since Menou et al. (\cite{mnl99})
have already noticed that although $f_t\approx 0.25$ for most
systems, V404 Cyg required $f_t\lta 0.1$. The value of $f_t$
should be predicted by the model describing the ``evaporation" of
an accretion disc into an ADAF, but we lack such a model.

\subsection{Variability}

The flux we obtained for GRO J1655-40 is consistent with that
found by Chandra (Kong et al. \cite{kmg02}), the difference being
of order of 50 \% for the unabsorbed flux in the 0.2 - 10 keV
range, and can be essentially accounted for by statistical
fluctuations; we do not find significant short time scale
variability, which means that these sources are relatively stable.
However, both XMM-Newton and Chandra flux measurements are one
order of magnitude below that of ASCA (Asai et al. \cite{adh98};
Ueda et al. \cite{uit98}). This observation, however, was
performed a month before the 1996 outburst and roughly 9 months
after the previous (August 1995) eruption. Clearly, the system was
not in its quiescent state during which it takes more than 30
years (the recurrence time) to refill the disc emptied by the
outburst. Matter was then transferred from the secondary at a rate
much higher than during the true quiescence (Esin et al.
\cite{elh00}).

Short timescale variability has been detected transients during
quiescence containing black holes, such as V404 Cyg (Wagner et al.
\cite{wsh94}; Kong et al. \cite{kmg02}), as well as those
containing neutron stars, such as Aql X-1 (Rutledge et al.
\cite{rbb02}). Such variability is not a surprise, but cannot be
significantly constrained in our data due to the low count rate.

\section{Conclusions}

We have observed and detected GRO J1655-40 during a 40 ksec
XMM-Newton observation, with a luminosity of 5.9 10$^{31}$ erg
s$^{-1}$; GRS 1009-45 was undetected, leading to a relatively
small upper limit on luminosity of 8.9 10$^{30}$ erg s$^{-1}$ for
a distance of 5 kpc. These observations are consistent with the
disc instability model if the accretion disc is truncated at a
radius of 0.1 - 0.3 times the circularization radius, where the
accretion flow forms an ADAF. The spectrum appears to be quite
hard, much harder than one would expect from coronal emission from
the secondary star or from the accretion disc itself. The quality
of the data is however not sufficient to constrain and distinguish
between various types of optically thin flows in the vicinity of
the black hole.

There is no sign of variability on a time scale of hours, and the
flux measured by XMM-Newton from GRO J1655-40 is consistent with
the Chandra value (Kong et al. \cite{kmg02}), and about 10 times
lower than that which was measured with ASCA by Asai et al.
(\cite{adh98}) and Ueda et al. (\cite{uit98}) between two
outbursts, indicating that the system was not fully in quiescence
at that time, and that the mass transfer rate from the secondary
was high.

\section*{Acknowledgments}

We are grateful to R. Narayan for discussions and comments on the
manuscript. This work was supported in part by a grant from the
Centre National d'Etudes Spatiales.

%\listofobjects


\begin{thebibliography}{}

\bibitem[1998]{adh98}Asai, K, Dotani, T., Hoshi, R., Tanaka, Y.,
Robinson, C.R., \& Terada, K., 1998, PASJ, 50, 611

\bibitem[2000]{betal00}Barret, D., Olive,
J.~F., Boirin, L., Done, C., Skinner, G.~K., \& Grindlay, J.~E.\
2000, ApJ, 533, 329

\bibitem[2002]{bp02}Beer, M.E., \& Podsiadlowski, P. 2002, MNRAS,
331, 351

\bibitem[2001]{br01}Bildsten L., Rutledge R.E., 2001, ApJ 541, 908

\bibitem[2001]{cps01}Campana, S., Parmar, A.N., \& Stella, L.
2001, A\&A, 372, 241

\bibitem[2002]{csg02}Campana, S., et al. 2002, ApJ, 575, L15

\bibitem[1979]{c79}Cash, W. 1979, ApJ, 228, 939

\bibitem[1997]{csl97}Chen, W, Schrader, C.R., \& Livio, M. 1997,
ApJ, 491, 312

\bibitem[1997]{dbc97}della Valle, M., Benetti, S.;, Cappellaro, E., \& Wheeler,
C. 1997, A\&A, 318, 179

\bibitem[2001]{dhl01}Dubus, G., Hameury, J.M., \& Lasota, J.P.
2001, A\&A, 373, 251

\bibitem[2000]{elh00}Esin, A.~A.,
Lasota, J.-P., \& Hynes, R.~I.\ 2000, A\&A, 354, 987

\bibitem[1989]{fgm89}Fleming,
T.~A., Gioia, I.~M., \& Maccacaro, T.\ 1989, ApJ, 340, 1011

\bibitem[2002]{fkr02}Frank, J., King,
A., \& Raine, D.\ 2002, Accretion power in astrophysics.~3rd ed.,
Cambridge University Press

\bibitem[2001]{gmn01}Garcia, M.R., McClintock, J.E., Narayan, R.,
Callanan, P., Barret, D., \& Murray, S.S. 2001, ApJ, 553, L47

\bibitem[1998]{gm98}Gammie, C.F., \& Menou, K. 1998, ApJ, 492, L75

\bibitem[1997]{hlmn97}Hameury, J.-M., Lasota, J.-P.,
McClintock, J.~E., \& Narayan, R.\ 1997, ApJ, 489, 234

\bibitem[1998]{hmd98}Hameury, J.-M., Menou, K., Dubus, G.,
Lasota, J.-P., \& Hur\'e, J.-M. 1998, MNRAS, 298, 1048

\bibitem[1998]{hhs98}Hynes, R. I., Haswell, C. A., Shrader, C. R., Chen, W., Horne, K.,
et al. 1998, MNRAS, 300, 64

\bibitem[1998]{kr98}King, A.~R.~\& Ritter,
H.\ 1998, MNRAS, 293, L42

\bibitem[2002]{kmg02}Kong, A.K.H., McClintock, J.E., Garcia,M.R.,
Murray, S.S., \& Barret, D. 2002, ApJ, 570, 277

\bibitem[1996]{l96}Lasota, J.-P. 1996, IAU
Symp.~165: Compact Stars in Binaries, 165, 43

\bibitem[2000]{l00}Lasota, J.-P. 2000, A\&A, 360, 575

\bibitem[2001]{l01}Lasota, J.-P. 2001, NewAR, 45, 449

\bibitem[2002]{l02}Lasota, J.-P. 2002, ASP
Conf.~Ser.~261: The Physics of Cataclysmic Variables and Related
Objects, 397

\bibitem[1998]{lh98}Lasota, J.-P.~\&
Hameury, J.-M.\ 1998, Accretion Processes in Astrophysical
Systems: Some Like it Hot!, 351

\bibitem[1996]{lny96}Lasota, J.-P.,
Narayan, R., \& Yi, I.\ 1996, A\&A, 314, 813

\bibitem[1995]{mhr95}McClintock, J.E., Horne, K., \& Remillard,
R.A. 1995, ApJ, 442, 358

\bibitem[2002]{m02}Menou, K., 2002, in The physics of
cataclysmic variables and related objects, ed. B.T. G{\"a}nzicke,
K. Beuermann \& K. Reinsch, ASP Conf. Ser. 261, p. 387

\bibitem[1999a]{mnl99}Menou, K.,
Narayan, R., \& Lasota, J.\ 1999, ApJ, 513, 811

\bibitem[1999b]{men99}Menou, K., Esin, A.A., Narayan, R., Garcia,
M.R., Lasota, J.P., \& McClintock, J.E. 1999, ApJ, 520, 276

\bibitem[2000]{mhln00}Menou, K., Hameury, J.M., Lasota, J.P., \&
Narayan, R. 2000, MNRAS, 314, 498

\bibitem[1996]{nmy96}Narayan, R., McClintock, J.E., \& Yi, I.,
1996, ApJ, 457, 821

\bibitem[1997a]{nbm97}Narayan, R., Barret, D., \& McClintock, J.E.
1997, ApJ, 482, 448

\bibitem[1997b]{ngm97}Narayan, R., Garcia, M.~R., \& McClintock,
J.~E.\ 1997, ApJ, 478, L79

\bibitem[2002]{ngm02}Narayan R., Garcia M.R., McClintock J.E., 2002,
Proc. IX Marcel Grossmann Meeting, eds. V. Gurzadyan, R. Jantzen
and R. Ruffini, Singapore: World Scientific, in press,
astro-ph/0107387

\bibitem[2001]{ns01}Nayakshin S., \& Svensson R., 2001, ApJ 551, L67

\bibitem[1999]{qn99} Quataert, E.~\&
Narayan, R.\ 1999, ApJ, 520, 298

\bibitem[2001]{rpm01}Ramsay, G., Poole, T., Mason, K., C\'ordova, F.,
Priedhorsky, W., et al. 2001, A\&A, 365, L288

\bibitem[2002]{rbb02}Rutledge, R.E., Bildsten, L., Brown, E.F.,
Pavlov, G.G., \& Zavlin, V.E. 2002, ApJ, 577, 346

\bibitem[1990]{scs90}Schmitt, J.H.M.M., Collura, A, Sciortino,
S., Vaiana, G.S., Hardnen Jr, F.R., Rosner, R., ApJ, 365, 704

\bibitem[2002]{skc02}Sutaria, F.K., Kolb, U., Charles, P., et al.
2002, A\&A, 391, 993

\bibitem[1996]{ts96}Tanaka, Y., \& Shibazaki, N. 1996, ARA\&A, 34, 607

\bibitem[1998]{uit98}Ueda, Y.,Inoue, H., Tanaka, Y., Ebisawa, K., Nagase, F., Kotani, T.,
\& Gehrels, N., 1998, ApJ, 492, 782

\bibitem[1994]{vbj94}Verbunt, F., Belloni, T., Johnston, H. M., van der Klis, M., \&  Lewin, W. H.
G.1994, A\&A, 285, 903

\bibitem[1994]{wsh94}Wagner, R.M., Starrfield, S.G., Hjellming, R.M., Howell, S.B., \& Kreidl, T.
J. 1994, ApJ, 429, L25

\bibitem[2002]{w02}Wijnands, R. 2002, ApJL, submitted
(Astro-ph/0207102)

\end{thebibliography}
\end{document}